\RequirePackage{fixltx2e}
\documentclass[aip,jcp,reprint,floatfix]{revtex4-1}

\usepackage{amsmath,amsfonts} %
\usepackage{wasysym} %
\usepackage[acronym]{glossaries} %
\usepackage{graphicx} %
\usepackage{dcolumn} %
\usepackage{bm} %
\usepackage{ifpdf} %
\usepackage{color} %
\usepackage{nicefrac} %
\usepackage{natbib} %
\usepackage[byname]{smartref} %
\usepackage[breaklinks, %
colorlinks, %
linkcolor=blue, %
citecolor=blue, %
urlcolor=blue]{hyperref} %
\usepackage[table]{xcolor} %
\usepackage{braket} %

\DeclareMathOperator\erf{erf} 
\DeclareMathOperator\diag{diag} 
\newcommand{\bfr}{\mathbf{r}} %
\newcommand{\eq}[1]{Eq.~(\ref{#1})} %
\newcommand{\bea}{\begin{eqnarray}}
\newcommand{\eea}{\end{eqnarray}}

\ifpdf %
 \fi

\begin{document}
                            
\newacronym{CI}{CI}{conical intersection} %
\newacronym{GP}{GP}{geometric phase} %
\newacronym{LVC}{LVC}{linear vibronic coupling} %
\newacronym{DOF}{DOF}{degrees of freedom} %
\newacronym{PES}{PES}{potential energy surface} %
\newacronym{DBOC}{DBOC}{diagonal Born--Oppenheimer correction} %
\newacronym{BMA}{BMA}{bis(methylene) adamantyl} %
\newacronym{FC}{FC}{Franck-Condon} %
\newacronym{CWE}{CWE}{cylindrical wave expansion}
\newacronym{DVBC}{DVBC}{double value boundary conditions}

\title{When do we need to account for the geometric phase in excited state dynamics?}

\author{Ilya G. Ryabinkin} %
\author{Lo{\"i}c Joubert-Doriol} %
\author{Artur F. Izmaylov} %
\affiliation{Department of Physical and Environmental Sciences,
  University of Toronto Scarborough, Toronto, Ontario, M1C 1A4,
  Canada} %
\affiliation{Chemical Physics Theory Group, Department of Chemistry,
  University of Toronto, Toronto, Ontario M5S 3H6, Canada} %

\date{\today}

\begin{abstract}
 We investigate the role of the \gls{GP} in an internal conversion process
  when the system changes its electronic state by passing through a
  \gls{CI}. Local analysis of a two-dimensional \gls{LVC}
  model Hamiltonian near the \gls{CI} shows that the role of the \gls{GP} is twofold. First, it
  compensates for a repulsion created by the so-called \gls{DBOC}. 
  Second, the \gls{GP} enhances the non-adiabatic transition probability for a wave-packet part that
  experiences a central collision with the \gls{CI}. To assess the significance of both 
  \gls{GP} contributions we propose two indicators that can be computed from parameters of electronic 
  surfaces and initial conditions. To generalize our analysis to $N$-dimensional systems
  we introduce a reduction of  a general $N$-dimensional LVC model
  to an effective 2D LVC model using a mode transformation that preserves short-time dynamics
  of the original $N$-dimensional model. 
  Using examples of the bis(methylene) adamantyl and butatriene cations, 
  and the pyrazine molecule we have demonstrated that their effective 2D models reproduce 
 the short-time dynamics of the corresponding full dimensional models, 
  and the introduced indicators are very reliable in assessing GP effects. 
  
\end{abstract}


\maketitle

\glsresetall

\section{Introduction}
\label{sec:introduction}

\Glspl{CI} of electronic states provide an efficient
mechanism for radiationless electronic transitions.~\cite{Koppel:1984/acp/59, Yarkony:1996/rmp/985,
  Hahn:2000/jpcb/1146} \glspl{CI} act as 
``funnels''~\cite{Balzer:2003/cpl/351} for the nuclear density and enables rapid
conversion of the excessive electronic energy into nuclear motion. 
Owing to the ubiquity of \glspl{CI} in molecules,~\cite{Cederbaum:1977/cp/169,
  Seidner:1993/cpl/117,Woywod:1994/jcp/1400, Kendrick:1996/jcp/7502,
  Hahn:2000/jpcb/1146, Cattarius:2001/jcp/2088,
  Vallet:2005/jcp/144307, Burghardt:2008/jcp/174104,
  Sardar:2008/pccp/6388, Sirjoosingh:2011/jctc/2831,
  Domcke:2012/arpc/325, Halasz:2012/jpca/2636, Ou:2013/jpcc/19839}
an adequate theoretical description of this conversion mechanism is an
important task in theoretical physical chemistry.

Conical intersections of potential energy surfaces lead not only to
non-adiabatic transitions but also to the appearance of the \gls{GP}~\cite{Berry:1984/rspa/45,
Mead:1979/jcp/2284, Berry:1987/rspa/31} in both electronic and nuclear wave-functions. 
The \gls{GP} manifests in a sign change of adiabatic electronic wave-functions 
along a closed path of nuclear configurations encircling the \gls{CI} seam.\cite{LonguetHigg:1958/rspa/1,Mead:1979/jcp/2284}
This sign change must be compensated by corresponding nuclear
wave-functions in order to preserve the singe-valued character of the total wave-function. 
The \gls{GP}  poses a challenge for modelling non-adiabatic dynamics 
because nuclear wave-functions must be simulated with \gls{DVBC}. 
Neglecting \gls{DVBC} for low energy nuclear dynamics 
on the ground electronic state near the \gls{CI} can result in qualitatively wrong predictions.\cite{Babikov:2005ja,Babikov:2010hm} 
The \gls{GP} causes an extra phase accumulation for fragments of the nuclear 
wave-packet that skirt the \gls{CI} on opposite sides.\cite{Schon:1995/jcp/9292,Ryabinkin:2013/prl/220406} 
Resulting destructive interference can lead either to a spontaneous localization of the nuclear
density~\cite{Ryabinkin:2013/prl/220406} or slower nuclear dynamics
\cite{Loic:2013/jcp/234103} than in the case where the \gls{GP}
is neglected. 

A question arises about the role of the \gls{GP} in the excited 
state nuclear dynamics through the CI.  
Recently, Althorpe and co-workers put forward a topological analysis considering
Feynman path integral trajectories and their ``winding numbers''.
\cite{JuanesMarcos:2005/sci/1227, Althorpe:2006/jcp/084105,Althorpe:2008/jcp/214117}
Practically, for photo-induced interconversion processes this analysis 
involves numerical simulation of quantum nuclear wave-packet 
dynamics with and without \gls{GP}-induced \gls{DVBC} and evaluation of wave-packet 
components that are even and odd with respect to the $2\pi$ rotation around the CI. 
The spatial overlap between the even and odd components provides a measure of GP significance. 
The obvious difficulty with this analysis is a necessity of quantum dynamics 
with and without \gls{GP}-induced \gls{DVBC}, such simulations cannot be easily done for a 
general molecular system.

Recent studies~\cite{Althorpe:2008/jcp/214117, Bouakline:2014/cp/} of
non-adiabatic transitions in photodissociation of pyrrole have shown that the
 impact of the \gls{GP} on the dynamics near $^1B_1$ -- $S_0$ and $^1A_2$ -- $S_0$
CIs is quite different. For $^1B_1$ -- $S_0$ it changed the branching ratio between 
two fragmentation products only slightly, while for $^1A_2$ -- $S_0$ 
stronger \gls{GP} effects were found. To date, no satisfactory explanation of this difference has been given. 
On the other hand, several research groups are actively develop  on-the-fly non-adiabatic
dynamics techniques following mixed quantum-classical approach that neglects GP effects. 
It is not clear how results of these techniques would change if GP effects were included.
Thus, it is highly desirable to build a theory that can predict the significance of the GP 
without performing full dimensional quantum nuclear dynamics simulations.

To address this challenge we begin our consideration with analysis of one 
of the simplest two-dimensional diabatic models that can provide the CI 
in the adiabatic representation.\cite{Jahn:1937/prca/220, LonguetHigg:1958/rspa/1,Bersuker:2001/cr/1067, Child:2003}
In the 2D model we demonstrate that for excited state dynamics 
mostly local properties in a vicinity of the CI define significance of the GP, and a great
body of system-specific information on a periphery of the CI is secondary. 
Considering an $N$-dimensional extension of our model we propose a transformation 
that reduces the system dimensionality back to two while preserving short-time 
dynamics of the $N$-dimensional case.  Using this reduction transformation 
we extend the 2D analysis to $N$-dimensional models.
Finally, from the local analysis we devise characteristics that can be 
obtained from electronic structure calculations for molecules and 
can predict significance of GP effects for molecular non-adiabatic dynamics.

Note that the GP appears only  in the adiabatic representation because it is a property 
of adiabatic electronic and nuclear wave-functions. In the diabatic 
representation~\cite{Smith:1963/pr/111,Baer:1975/cpl/112, Mead:1982/jcp/6090} 
the GP is absent. Still, due to equivalence of the diabatic and adiabatic representations, 
dynamical features that appear only when the GP is included in the adiabatic representation 
are present in the diabatic dynamics but constitute its indiscernible from other effects part. 
Thus, due to absence of the GP in the diabatic representation, it is 
easier numerically to perform exact dynamics in that representation.
The main problem with the diabatic representation is that it cannot be rigorously 
defined for a finite number of electronic states in a general molecular 
system.~\cite{Baer:1975/cpl/112, Mead:1982/jcp/6090} 
The adiabatic representation is a primary representation available from the first-principles
(\textit{ab initio}) calculations for molecules, and the diabatic representation is usually 
obtained from the adiabatic representation in some approximate way.
\cite{Koppel:2006/mp/1069,Voorhis:2010/arpc/149,Sirjoosingh:2011/jctc/2831,
  Yarkony:2012/cr/481,Subotnik:2009/jcp/234102} 
  However, since the reversed transformation from the
diabatic to the adiabatic representation is always exact, we use diabatic models and the associated 
adiabatic representation to analyze GP effects.  

This paper is organized as follows. First, by analyzing the
difference between model Hamiltonians with and without account for the \gls{GP} 
we identify two main \gls{GP} effects that modify
non-adiabatic dynamics. Second, we discuss two
indicators that allow us to assess the importance of \gls{GP} effects without simulating quantum dynamics. 
Third, we simulate and analyze non-adiabatic dynamics for a few molecular systems that provide 
a variety of dynamical regimes and allows us to probe 
limitations of our theoretical analysis.
Finally, we conclude the paper with a summary and an outlook for future work.
Atomic units are used throughout this paper.

\section{Theoretical analysis}
\label{sec:theoretical-analysis}

\subsection{Two-dimensional linear vibronic coupling model}
\label{sec:two-dimens-line}

We begin our consideration of the two-dimensional \gls{LVC} model with
its Hamiltonian
\begin{equation}
  \label{eq:H_lvc}
  \hat H = {\hat T}  {\mathbf 1}_2 + 
  \begin{pmatrix} 
    V_{11} & V_{12} \\
    V_{12} & V_{22}
  \end{pmatrix},
\end{equation}
where $\hat T = -\frac{1}{2}\nabla^2 \equiv -\frac{1}{2}(\partial^2
/\partial x^2 +\partial^2 /\partial y^2) $ is the nuclear kinetic
energy operator, and ${\mathbf 1}_2$ is a $2\times 2$ unit matrix.  $ V_{11}$ and
$ V_{22}$ are the diabatic potentials represented by identical 2D
parabolas shifted in the $x$-direction by $a$, in energy by $\Delta$
\begin{align}
  \label{eq:diab-me-11}
  V_{11} = {} & \frac{\omega_1^2}{2}\left(x + \frac{a}{2}\right)^2
  + \frac{\omega_2^2}{2}y^2 + \frac{\Delta}{2},\\
  \label{eq:diab-me-22}
  V_{22} = {} & \frac{\omega_1^2}{2}\left(x - \frac{a}{2}\right)^2 +
  \frac{\omega_2^2}{2}y^2 - \frac{\Delta}{2}.
\end{align}
To have the \gls{CI} in the adiabatic representation $ V_{11}$ and
$ V_{22}$ are coupled by the linear $ V_{12}=c y$ potential.
Switching to the adiabatic representation for the 2D LVC Hamiltonian in \eq{eq:H_lvc} is done by diagonalizing the potential
matrix using a unitary transformation 
\begin{equation}
  U = 
  \label{eq:Umat}
  \begin{pmatrix}
    \cos\theta & \sin\theta \\
    -\sin\theta & \cos\theta
  \end{pmatrix},
\end{equation}
where $\theta$ is a mixing angle between the diabatic electronic
states states $\ket{1}$ and $\ket{2}$
\begin{equation}
  \label{eq:theta}
  \theta = \frac{1}{2}\arctan \dfrac{2\,V_{12}}{V_{11} - V_{22}} =
  \frac{1}{2}\arctan \dfrac{\gamma y}{x + b}.
\end{equation}
Here, $b = \Delta/(\omega_1^2 a)$ is the $x$-coordinate of the CI
point, and $\gamma = {2c}/{(\omega_1^2a)}$ is dimensionless coupling
strength. For simplicity of the subsequent analysis we set $b =0$,
which corresponds to centering the coordinates at the \gls{CI} point.

\begin{figure*}
  \centering
  \includegraphics[width=\textwidth]{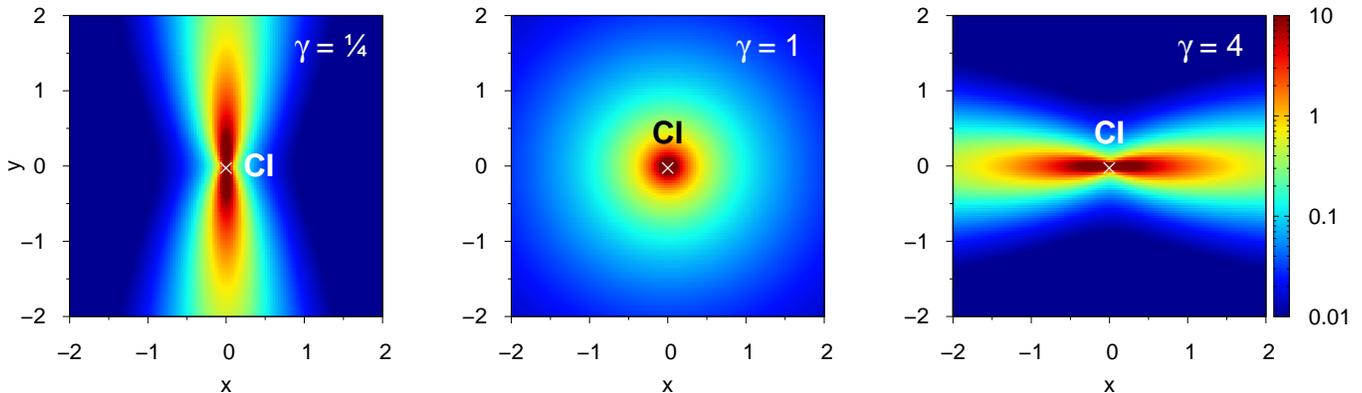}
  \caption{The diagonal Born-Oppenheimer correction,
    Eq.~\eqref{eq:tau-adi-diag}, for different values of $\gamma$.}
  \label{fig:dboc}
\end{figure*}

The transformation in \eq{eq:Umat} gives rise to the 2D \gls{LVC}
Hamiltonian in the adiabatic representation $\hat H_\text{adi}  = U
{\hat H} U^\dagger$,
\begin{equation}
  \label{eq:adiab}
   \hat H_\text{adi}  =   
  \begin{pmatrix}
    \hat T + \hat\tau_{11}& i\hat\tau_{12} \\
    -i\hat\tau_{21} & \hat T +\hat\tau_{22}
  \end{pmatrix} +
  \begin{pmatrix}
    W_{-} & 0 \\
    0 & W_{+}
  \end{pmatrix},
\end{equation}
where
\begin{align}
  \label{eq:Wmin}
  W_{\pm} = & {} \dfrac{1}{2}\left(V_{11} + V_{22}\right) \pm
  \dfrac{1}{2}\sqrt{\left(V_{11} - V_{22}\right)^2 + 4 V_{12}^2},
\end{align}
are the adiabatic potentials and $\hat\tau_{ij}$ are the non-adiabatic
couplings. For our model we can further express $\hat\tau_{ij}$ as
\begin{align}
  \label{eq:tau-adi-diag}
  \hat\tau_{11} & {} =\hat\tau_{22} = \frac{1}{2}
  \nabla\theta\cdot\nabla\theta =
  \frac{x^2 + y^2}{8(\gamma^{-1}x^2 +\gamma y^2)^2}, \\
  \hat\tau_{12} & {} = \hat\tau_{21} =
  \frac{i}{2}\left(\overleftarrow\nabla\cdot\nabla\theta-\nabla\theta\cdot\overrightarrow\nabla\right)  \nonumber  \\
  \label{eq:tau-adi-offd}
  & {} = \frac{(\overrightarrow L_z-\overleftarrow {L_z})}{4(\gamma^{-1}x^2 +\gamma y^2)},
\end{align}
where $L_z = x p_y - y p_x$ is the $z$ component of the angular
momentum operator, and the overhead arrows indicate 
the directions in which the differential operators act.\footnote{The direction of action of $L_z$
must be specified because $L_z$ does not commute with the denominator $(x^2 +
  \gamma^2y^2)$ unless $\gamma = 1$.} The diagonal non-adiabatic
couplings, $\hat\tau_{11}$ and $\hat\tau_{22}$, represent a repulsive
potential known as the \gls{DBOC}.\cite{Handy:1996/cpl/425,
  Valeev:2003/jcp/3921} The DBOC is the parametric function of the coupling strength
parameter $\gamma$. Figure~\ref{fig:dboc} illustrates the DBOC for 
representative values of $\gamma$. The off-diagonal elements, $\hat\tau_{12}$ and
$\hat\tau_{21}$ in Eq.~\eqref{eq:tau-adi-offd}, couple dynamics on 
the adiabatic potentials $W_{\pm}$ and are responsible for 
non-adiabatic transitions. 
 
 If we simulate the spectrum or nuclear dynamics for $\hat H_\text{adi}$ using  
 single-valued basis functions, the outcome can be very different from that for 
 the original full Hamiltonian $\hat H$.\cite{Ryabinkin:2013/prl/220406} This difference 
 arises as a result of ignoring proper \gls{DVBC} for the $\hat H_\text{adi}$
 Hamiltonian. The unitary transformation $U$ changes its sign if one encircles 
 the CI point using the parametric dependence of $\theta$ on the nuclear coordinates $x$ and $y$. 
 Since the adiabatic electronic states are columns of the $U$ matrix in the diabatic basis, 
 this sign change is a manifestation of the \gls{GP} that is acquired by the adiabatic 
 electronic wave-functions.\cite{Berry:1984/rspa/45,
Mead:1979/jcp/2284, Berry:1987/rspa/31}
The total electron-nuclear wave-function is always single-valued and thus \gls{DVBC} 
for the electronic part impose the \gls{DVBC} for the corresponding nuclear part. 
As a consequence, to simulate the nuclear dynamics of $\hat H_\text{adi}$ 
with a proper account of \gls{GP} effects, one needs to impose \gls{DVBC}. 
To assess the importance of GP effects we use simulations of the $\hat H_\text{adi}$ non-adiabatic dynamics
\emph{without} imposing \gls{DVBC} as a reference which will be referred to as the ``no GP'' model.

To account for GP effects in the adiabatic representation we 
follow the~\citet{Mead:1979/jcp/2284} technique that 
introduces a position-dependent phase
factor $e^{i\theta}$ with $\theta$ given by Eq.~\eqref{eq:theta}.
This phase factor changes the sign upon encircling the \gls{CI} and 
can be either attached to nuclear basis functions 
to introduce \gls{DVBC} or alternatively used to transform  $\hat H_\text{adi}$ into
$\hat H_\text{GP} = e^{-i\theta} \hat H_\text{adi}  e^{i\theta}$. We will follow the second path 
because GP effects in the $\hat H_\text{GP}$ Hamiltonian have a concrete operator 
representation that will facilitate our analysis. Since the $e^{i\theta}$ transformation of 
$\hat H_\text{adi}$ contains only functions of nuclear coordinates, $\hat H_\text{GP}$ 
is different from $\hat H_\text{adi}$ only in the kinetic energy part
\bea
   \hat H_\text{GP}  =   
  \begin{pmatrix}
    \hat T + \hat\tau_{11}^{\rm GP}& i\hat\tau_{12}^{\rm GP} \\
    -i\hat\tau_{21}^{\rm GP} & \hat T +\hat\tau_{22}^{\rm GP}
  \end{pmatrix} +
  \begin{pmatrix}
    W_{-} & 0 \\
    0 & W_{+}
  \end{pmatrix},
\eea
where
\begin{align}
 \label{eq:tau-GP-diag}
 \hat\tau_{11}^\text{GP} & {} = \hat\tau_{22}^\text{GP} = \hat\tau_{11} +
 \left(e^{-i\theta}
   \hat T e^{i\theta} - \hat T \right) \nonumber \\
 & {} = \hat\tau_{11} + \frac{i}{2}\left(\overleftarrow\nabla\cdot\nabla\theta -\nabla\theta\cdot\overrightarrow\nabla \right) +
 \frac{1}{2}\nabla\theta \cdot \nabla\theta \nonumber \\
 & {} = \frac{(\overrightarrow L_z-\overleftarrow {L_z})}{4(\gamma^{-1}x^2 +\gamma y^2)} +
 \frac{x^2 + y^2}{4(\gamma^{-1}x^2 +\gamma y^2)^2}, \\
 \hat\tau_{12}^\text{GP} & {} = \hat\tau_{21}^\text{GP} = e^{-i\theta}
 \hat\tau_{12}
 e^{i\theta}  \nonumber \\
 \label{eq:tau-GP-offd}
 & = \frac{(\overrightarrow{L_z}-\overleftarrow {L_z})}{4(\gamma^{-1}x^2 +\gamma y^2)} - \frac{x^2 +
   y^2}{4(\gamma^{-1}x^2 +\gamma y^2)^2}.
\end{align}
Comparison of Eqs.~\eqref{eq:tau-GP-diag}--\eqref{eq:tau-GP-offd}
with Eqs.~\eqref{eq:tau-adi-diag}--\eqref{eq:tau-adi-offd}
shows that the \gls{GP} modifies the non-adiabatic coupling matrix
elements and thus changes probabilities of non-adiabatic
transitions.

\subsection{The role of the geometric phase in non-adiabatic
  transitions}
\label{sec:role-geometric-phase}

Below we further separate individual physical mechanisms
that stem from the mathematical differences in $\hat\tau$ operators 
for models with and without the \gls{GP}.

\subsubsection{Compensation of the DBOC repulsion}
\label{sec:dboc-repulsion}

The non-adiabatic couplings $\hat\tau_{ij}$ and $\hat\tau_{ij}^\text{GP}$ are
singular at the \gls{CI}, therefore we can neglect any regular
operator in a vicinity of the \gls{CI}. In particular, the difference
of the adiabatic potentials is not only a regular operator but also
vanishes at the \gls{CI} [$W_{+}(\mathbf{r}_{\rm CI}) = W_{-}(\mathbf{r}_{\rm CI})$].  
This allows us to consider the kinetic matrix in Eq.~\eqref{eq:adiab} alone.
Applying a unitary rotation in the electronic subspace we can diagonalize the kinetic matrix 
\bea
 \frac{1}{2}\begin{pmatrix}
    1& i\\
    1 & -i
  \end{pmatrix}
  \begin{pmatrix}
    \hat T + \hat\tau_{11}& i\hat\tau_{12} \\
    -i\hat\tau_{21} & \hat T +\hat\tau_{22}
  \end{pmatrix}
    \begin{pmatrix}
    1& 1\\
    -i & i
  \end{pmatrix} \\ \notag
  = 
   \begin{pmatrix}
    \hat T + \hat\tau_{-}& 0\\
    0 & \hat T +\hat\tau_{+}
  \end{pmatrix},
\eea
where $\hat\tau_{\pm} = \hat\tau_{11} \pm \hat\tau_{12}$ (note that $\hat\tau_{11} =
\hat\tau_{22}$ and $\hat\tau_{12} = \hat\tau_{21}$). Eigenstates 
of decoupled sub problems
\begin{equation}
  \label{eq:gen-eigp}
  (\hat T + \hat\tau_{\pm})\chi_n^{\pm} = \epsilon_i^{\pm}\chi_i^{\pm},
  \quad i=1,\ldots  
\end{equation}
represent a convenient complete set of functions to expand any
wave-function in a vicinity of the \gls{CI}. However, properties of
the eigenstates $\chi_i^{\pm}$ are quite different for the models with
and without the \gls{GP}. Using Eqs.~\eqref{eq:tau-adi-diag} and
\eqref{eq:tau-adi-offd}, we can write $\hat\tau_{\pm}$ for the ``no \gls{GP}'' case as
\bea
\hat\tau_{\pm} = \frac{(x^2 + y^2)\pm(\overrightarrow L_z -\overleftarrow {L_z} )}{8(\gamma^{-1}x^2 +\gamma y^2)^2},
\eea
where both coupling and DBOC terms are intermixed. 
The singular \gls{DBOC} term gives rise to a cusp behavior at the \gls{CI} point for the eigenstates $\chi_i^{\pm}$ 
to maintain finite energy. A cusp-less wave-packet of finite energy cannot reach the \gls{CI},
since any finite-energy expansion in terms of the eigenstates $\chi_i^{\pm}$ must have a node at the \gls{CI}.
In contrast, with the \gls{GP} [Eqs.~\eqref{eq:tau-GP-diag} and \eqref{eq:tau-GP-offd}] we have
\bea
 \hat\tau_{+}^\text{GP}& = &\frac{\overrightarrow L_z -\overleftarrow {L_z}}{2(\gamma^{-1}x^2 +\gamma y^2)^2}, \\
 \hat\tau_{-}^\text{GP}& = &\frac{x^2 + y^2}{2(\gamma^{-1}x^2 +\gamma y^2)^2},
\eea
where there is a clear separation on the operator term $\hat\tau_{+}^\text{GP}$ 
and the scaled DBOC term $ \hat\tau_{-}^\text{GP}$.
Since $\hat\tau_{+}^\text{GP}$ does not contain the \gls{DBOC},
functions that satisfy $L_z \chi_i^{+} = 0$ form a suitable subset of
the eigenstates which can be \emph{finite} at the \gls{CI}. 
Therefore, in the presence of the \gls{GP} a cusp-less wave-packet can
access the \gls{CI} point.

Interpreting the \gls{DBOC} as physical repulsion, we can say
that this repulsion does not allow a wave-packet to reach the \gls{CI}
in the ``no GP'' case, whereas in the presence of the \gls{GP} this
repulsion is \emph{compensated} and the wave-packet can reach the
\gls{CI}. This compensation is especially important in the
small coupling case, $\gamma \ll 1$ (see Fig.~\ref{fig:dboc} $\gamma=1/4$), because 
the \gls{DBOC} represents a repulsive wall that can block
all parts of an incoming wave-packet from accessing regions
where the off-diagonal couplings, Eq.~\eqref{eq:tau-adi-offd}, are
large. If $\gamma \approx 1$  (see Fig.~\ref{fig:dboc} $\gamma=1$), the \gls{DBOC} 
repulsion and its GP compensation become less important because only a central 
part of the wave-packet is significantly repelled while 
peripheral parts can reach large coupling areas.

The idea of compensation of the \gls{DBOC} repulsion by the \gls{GP} can be further 
explored by introducing 
a simplified ``no \gls{GP}, no \gls{DBOC}'' Hamiltonian without the diagonal non-adiabatic
terms $\hat\tau_{ii}$
\begin{equation}
  \label{eq:Hsim}
  \hat H_{\rm adi}^\text{(s)} =   
  \begin{pmatrix}
    \hat T & i\hat\tau_{12} \\
    -i\hat\tau_{21} & \hat T
  \end{pmatrix} +
  \begin{pmatrix}
    W_{-} & 0 \\
    0 & W_{+}
  \end{pmatrix}.
\end{equation}
If the compensation mechanism is significant, the dynamics produced by 
$\hat H_{\rm adi}^\text{(s)}$ will be closer to that of the full Hamiltonian [\eq{eq:H_lvc}] than
to the ``no GP'' Hamiltonian [\eq{eq:adiab}].  

\subsubsection{Non-adiabatic transfer enhancement}
\label{sec:scattering-analysis}

The non-adiabatic couplings $\hat\tau_{ij}$ in models with or without the \gls{GP} 
contain the $z$-component of the angular momentum operator $L_z$ 
[see Eqs.~\eqref{eq:tau-adi-offd} and \eqref{eq:tau-GP-offd}]. 
Although the 2D LVC has cylindrical symmetry only when $\gamma = 1$ (see Fig.~\ref{fig:dboc}),
the presence of $L_z$ suggests to analyze dynamics of a wave-packet $\psi$ by expanding it in the
eigenstates of $L_z$
\begin{equation}
  \label{eq:wp-exp}
  \psi(x,y,t) = \sum_{m=-\infty}^\infty C_m(r, t) e^{-im\phi},
\end{equation}
where $r$ and $\phi$ are the polar coordinates centered at the
\gls{CI}. Action of the coupling on $\psi$ can be analyzed starting 
from the $L_z$ operator 
\begin{equation}
  \label{eq:Lz_wp_act}
  L_z \psi(x,y,t) = \sum_{m=-\infty}^\infty mC_m(r, t) e^{-im\phi}.
\end{equation}
Considering non-adiabatic transition for the $m=0$ component of $\psi$ we have
\bea
\hat\tau_{12}C_0(r,t) &=& \frac{(\overrightarrow L_z-\overleftarrow {L_z})}
   {4(\gamma^{-1}x^2 +\gamma y^2)} C_0(r,t) \\ \label{eq:Lz}
    &=&  \frac{-\overleftarrow {L_z} C_0(r,t)}{4(\gamma^{-1}x^2 +\gamma y^2)}. 
\eea
Introducing a resolution of the identity in the angular coordinate
\bea
\mathbf{1}_{\phi} = \frac{1}{2\pi}\int_0^{2\pi} d\phi \sum_{m'=-\infty}^{+\infty} e^{im'(\phi'-\phi)}
\eea
into Eq.~\eqref{eq:Lz} we obtain
 \bea\notag
\hat\tau_{12}C_0(r,t) &=& \frac{1}{2\pi}\int_0^{2\pi} d\phi \sum_{m'=-\infty}^{+\infty} e^{im'(\phi'-\phi)} \\
&&\times \frac{-\overleftarrow {L_z} C_0(r,t)}{4(\gamma^{-1}x^2 +\gamma y^2)}  \\ \notag
&=& \frac{1}{2\pi}\sum_{m'=-\infty}^{+\infty} e^{im'\phi'} m' C_0(r,t) \\ \label{eq:t12}
&& \times \int_0^{2\pi}  \frac{e^{-im'\phi}d\phi}{4[\gamma r^2 +(\gamma^{-1}-\gamma)r^2 \cos^2\phi]}.
 \eea 
If $\gamma=1$, the angular integral in \eq{eq:t12} becomes zero and there is no transfer for the $m=0$ 
component in the cylindrical symmetric case. For $\gamma\ne 1$, the angular integral in \eq{eq:t12}
is non-zero, and only $m'\ne 0$ contributions survive in the sum over $m'$. Therefore, even 
if cylindrical symmetry is broken, $\hat\tau_{12}$ can transfer the $m=0$ component only to the $m'\ne 0$ components. 
Due to an increase in kinetic energy associated with this process, 
the transfer probability is reduced compared to that for the $m\ne 0$ components of the initial wave-packet.    
 
For the model with \gls{GP}, $\hat\tau_{12}^\text{GP}$ in Eq.~\eqref{eq:tau-GP-offd} contains the
$L_z$-independent contribution. Thus, when $\gamma\ne 1$ the $m=0$ component can be 
transferred  into both $m'\ne 0$ and $m' =0$ components 
\bea
\hat\tau_{12}^{\rm GP}C_0(r,t) &=& \frac{-\overleftarrow {L_z} C_0(r,t)}{4(\gamma^{-1}x^2 +\gamma y^2)} 
- \frac{r^2 C_0(r,t)}{4(\gamma^{-1}x^2 +\gamma y^2)} \\ \notag
&=&  \frac{1}{2\pi}\sum_{m'=-\infty}^{+\infty} e^{im'\phi'} (m' -r^2)C_0(r,t) \\ \label{eq:t12gp}
&& \times \int_0^{2\pi}  \frac{e^{-im'\phi}d\phi}{4[\gamma r^2 +(\gamma^{-1}-\gamma)r^2 \cos^2\phi]}.
\eea
Opening the $m=0$ to $m'=0$ channel enhances the $m=0$ component transfer in the presence of the \gls{GP}. 
If the $m=0$ component dominates in the \gls{CWE} \eq{eq:wp-exp}, including the \gls{GP} 
will significantly alter non-adiabatic dynamics.

To estimate the significance of the \gls{GP} effect due to the $m=0$ transfer enhancement
we compute the \gls{CWE} at the moment $t_{\rm CI}$ of the closest proximity of 
a wave-packet to the \gls{CI} point. Once the coefficient $C_0(r,t_{\rm CI})$ is found, 
we evaluate the average weight of the $m = 0$ component as
\begin{equation}
  \label{eq:barw_ex}
  \bar w = \int r|C_0(r, t_{\rm CI})|^2\,dr.
\end{equation}
If $\bar w$ is much smaller than 50\%, the $m = 0$ component
is not dominant and including the \gls{GP} will not produce significant change
in nuclear dynamics.

Although $\bar w$ contains all necessary information about the $m=0$
component, dynamical simulations are required to compute it. However, 
for cases when the energy splitting (tuning) coordinate is strictly orthogonal 
to the coupling coordinate we can devise a simpler characteristic to 
assess the importance of the $m=0$ GP effect without running simulations.  
For that we resort to a semiclassical consideration assuming a frozen 
Gaussian form of the nuclear wave function. Due to the orthogonality of the tuning and coupling coordinates
the \gls{FC} point is shifted along the $x$ coordinate. Hence the nuclear 
wave function will have a momentum $\mathbf{p}=(p_x,0)$ upon arrival at the CI and the form
\begin{equation}
  \label{eq:Gaussian_wp}
  \Psi(x,y) = \sqrt{\frac{2}{\pi \sigma_x \sigma_y}}
  \exp{\left(-\frac{x^2}{\sigma_x^2} -
      \frac{y^2}{\sigma_y^2}\right)}\exp{(-ip_xx)}. 
\end{equation}
Considering $|\Psi(x,y)|^2$ as the density of an ensemble of classical
particles, each particle of this ensemble has an absolute value of the classical angular
momentum $|l_z| = |p_xy|$ with respect to the \gls{CI} point.
Using the relation between the momentum and kinetic energy $p_x^2/2 = E_\text{kin}$,
and the estimate of $E_\text{kin}$ as the difference between the potential energies of the wave-packet in 
 the initial position $W_{+}(\bfr_\text{ini}) $ and in the \gls{CI} point $W_{+}(\bfr_\text{CI})$, we have
\begin{equation}
  \label{eq:p-def}
  p_x = \sqrt{2\left[W_{+}(\bfr_\text{ini}) - W_{+}(\bfr_\text{CI})
    \right]}.  
\end{equation}
The angular momentum quantum number of the classical particle can be
estimated as $m \approx |p_xy|$, so that a region $-r_\text{eff} < y < r_\text{eff}$, where
\begin{equation}
  \label{eq:reff}
  r_\text{eff} = p_x^{-1}.
\end{equation}
corresponds to $|p_xy|<1$ values  and is assigned to the quantum value $m = 0$. 
Therefore, the average weight of the $m = 0$ component in
Eq.~\eqref{eq:barw_ex} can be approximated as
\bea   \notag
\bar w \approx \bar w_{\rm app} &=& 
  \int_{-\infty}^{\infty}dx\int_{-r_\text{eff}}^{r_\text{eff}}
  |\Psi(x,y)|^2\,dy \\
  &=& \erf{\left(\frac{\sqrt{2}r_\text{eff}}{\sigma_x}\right)} \\
  &=& \erf{\left(\frac{1}{\sigma_x\sqrt{W_{+}(\bfr_\text{ini}) - W_{+}(\bfr_\text{CI})}}\right)}.\label{eq:Cm_from_reff} 
\eea
Thus, evaluation of $\bar w_{\rm app}$ in \eq{eq:Cm_from_reff} does not require 
dynamical simulations and uses only the $\sigma_x$ parameter of the initial Gaussian 
and the adiabatic potential values $W_{+}(\bfr_\text{ini})$ and $W_{+}(\bfr_\text{CI})$.

\subsection{Extension to $N$-dimensional LVC model}
\label{sec:n-dimensional-case}

To extend our analysis to realistic molecular models we consider a general
$N$-dimensional linear vibronic coupling
model~\cite{Koppel:1984/acp/59} with the Hamiltonian
\begin{align}
  \label{eq:nd-lvc}
 \hat H_\text{ND} = & \sum_j^N \frac{1}{2}\left(p_j^2 + \omega_j^2 q_j^2
  \right)\mathbf{1}_2 +
  \begin{pmatrix}
    \kappa_j q_j & c_j      q_j \\
    c_j q_j & \tilde\kappa_j q_j
  \end{pmatrix}\nonumber\\
  &\hspace{1cm}+\begin{pmatrix}
    -\delta/2 & 0 \\
    0 & \delta/2
  \end{pmatrix},
\end{align}
where $q_j$ and $p_j$ are mass-weighted coordinates and conjugated
momenta, $\omega_j$ are frequencies, $\kappa_j$,$\tilde\kappa_j$, 
and $c_j$ are linear coupling constants, and 
$\delta$ is the energy difference between the two diabatic electronic 
potentials in the \gls{FC} point. As shown in previous studies
\cite{Meng:2013/jcp/014313, Li:2013/jcp/094313,Leveque:2013/jcp/044320} 
the $N$-dimensional LVC model can adequately reproduce vibronic spectra of molecular systems with \glspl{CI}.  
Another advantage of the $N$-dimensional LVC model is that its short time dynamics can be obtained from 
effective Hamiltonians of a lower dimensionality.\cite{Cederbaum:2005/prl/113003, Burghardt:2006/mp/1081,
  Gindensperger:2006/jcp/144103} There exist unitary transformations that rotate 
nuclear coordinates of $H_\text{ND}$ so that after a truncation of all but a few collective 
DOF essential dynamical characteristics of $H_\text{ND}$
(e.g., auto-correlation functions, electronic populations) can still be reproduced. 
In this work we use a reduction procedure that is similar in spirit to those used in 
Refs.~\onlinecite{Cederbaum:2005/prl/113003, Burghardt:2006/mp/1081,
Gindensperger:2006/jcp/144103,Loic:2013/jcp/234103} but different in its
focus on recovering a few lowest time derivatives 
of diabatic electronic populations of the full $N$-dimensional LVC Hamiltonian by its reduced counterpart.  
After the unitary rotation and truncation detailed in the Appendix we obtain the following effective 2D Hamiltonian
\begin{align}
 \nonumber
  \hat H_{2D} = {} & \left(\frac{P_X^2 + P_Y^2}{2} + \frac{{
        \Omega}_1^2 X^2 + {\Omega}_2^2 Y^2}{2}\right) \mathbf{1}_2 
          +
  \begin{pmatrix}
    \frac{1}{2}\Delta & \Delta_{12} \\
    \Delta_{12} & -\frac{1}{2}\Delta
  \end{pmatrix}\\ \label{eq:Halmost2D}
& {} +
  \begin{pmatrix}
    D_1 X + D_2 Y &  C_1 X + C_2 Y \\
    C_1 X + C_2 Y & -D_1 X - D_2 Y
  \end{pmatrix},
\end{align}
where $X,Y$ and $P_X,P_Y$ are collective coordinates and corresponding momenta, 
and $D_i, C_i, \Omega_i, \Delta,\Delta_{12}$ are constants defined in the Appendix. 
The Hamiltonian  $\hat H_{2D}$ [\eq{eq:Halmost2D}] can be seen as the generalization of the 
2D \gls{LVC} Hamiltonian. 
Due to molecular symmetry all systems studied in this work have 
$C_1 =D_2 =\Delta_{12}=0$ and thus, the 2D 
consideration can be extended directly to short-term dynamics of such $N$-dimensional systems.

\section{Numerical examples}
\label{sec:numerical-examples}

\begin{figure}
  \centering
  \includegraphics[width=0.5\textwidth]{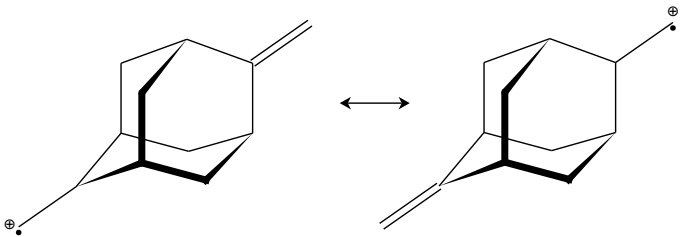}
  \caption{The bis(methylene) adamantyl cation has two 
    charge-localized conformation that are the result of a Jahn-Teller distortion from a 
    symmetric configuration of the CI seam minimum.}
  \label{fig:bma_iso}
\end{figure}

Here we consider three molecular systems with \glspl{CI}
 that are well described by multi-dimentional \gls{LVC} models: the
\gls{BMA}~\cite{Blancafort:2005/jacs/3391}(Fig.~\ref{fig:bma_iso}) 
and butatriene~\cite{Cederbaum:1977/cp/169, Koppel:1984/acp/59,
  Cattarius:2001/jcp/2088, Burghardt:2006/mp/1081,
  Gindensperger:2006/jcp/144104, Sardar:2008/pccp/6388} cations, and the
pyrazine molecule.~\cite{Seidner:1993/cpl/117, Woywod:1994/jcp/1400,
  Sukharev:2005/pra/012509, Burghardt:2008/jcp/174104} 
$N$-dimensional \gls{LVC} models for these systems are taken 
from literature\cite{Izmaylov:2011/jcp/234106,Cattarius:2001/jcp/2088,Raab:1999/jcp/936}. 
Our dimensionality reduction procedure is applied
to obtain parameters of 2D effective LVC Hamiltonians \eq{eq:H_lvc} (see Table~\ref{tab:BMA-param}).
\begin{table}
  \caption{Parameters of the 2D effective \protect\gls{LVC}
    Hamiltonian, Eq.~\eqref{eq:H_lvc}, for the studied systems.} 
  \label{tab:BMA-param}
  \centering
  \begin{ruledtabular}
    \begin{tabular}{@{}lcccr@{}}
      \multicolumn{1}{c}{$\omega_1$} & $\omega_2$ & $a$ & $c$ &
      \multicolumn{1}{c}{$\Delta$} \\ \hline
      \multicolumn{5}{c}{ Bis(methylene) adamantyl cation} \\
      $7.743\times10^{-3}$ & $6.680\times10^{-3}$ & 31.05 &
      $8.092\times 10^{-5}$ & 0.00000 \\[1ex]
      \multicolumn{5}{c}{ Butatriene cation} \\
      $9.557\times10^{-3}$ & $3.3515\times10^{-3}$  & 20.07   &
      $6.127\times 10^{-4}$ & 0.01984  \\[1ex]
      \multicolumn{5}{c}{ Pyrazine} \\
      $3.650\times10^{-3}$ & $4.186\times10^{-3}$ & 48.45 & $4.946\times
      10^{-4}$ & 0.02757 
    \end{tabular}
  \end{ruledtabular}
\end{table}
To quantify \gls{GP} effects we solve the time-dependent
Schr\"odinger equation in a finite basis for three model Hamiltonians derived from the 
effective 2D LVC Hamiltonian:
1) the full Hamiltonian [Eq.~\eqref{eq:H_lvc}], 2) the ``no GP'' Hamiltonian 
[Eq.~\eqref{eq:adiab}], and 3) the ``no GP, no DBOC'' Hamiltonian [Eq.~\eqref{eq:Hsim}].
For all three models we compare the adiabatic population dynamics 
$P_{\rm adi}(t)  =  \braket{\psi_\text{adi}^{(e)}(t)|\psi_\text{adi}^{(e)}(t)}$,
where $\psi_\text{adi}^{(e)}(x,y,t)$ is a time-dependent nuclear wave-function
that corresponds to the excited adiabatic electronic state. 
There are two sets of the initial conditions employed in this work: i) a wave-packet 
is taken as a Gaussian function \eq{eq:Gaussian_wp} 
with widths $\sigma_x =\sqrt{2/\omega_1}$ and $\sigma_y = \sqrt{2/\omega_2}$ and ii) the same Gaussian 
function as in (i) but multiplied by the $y$ coordinate. 
In both sets the initial position of a wave-packet is chosen at the \gls{FC} point
of the ground state of the corresponding full-dimensional models, and the initial momentum 
is set to zero.
If the first set of initial conditions corresponds to a regular setup of an ultrafast laser photo experiment, 
the second setup has been designed to assess the importance of the GP effect associated with non-adiabatic transfer 
of the $m=0$ component.  Multiplication of the Gaussian function by $y$ creates 
the nodal line $y = 0$ in the wave-packet and eliminates the 
$m = 0$ component from the corresponding \gls{CWE}~\eqref{eq:wp-exp}. 

To connect the results of our numerical calculations to our theoretical analysis in Table~\ref{tab:GP-param} we present 
parameters  that are most relevant to GP effects for all studied systems. Among other parameters
we found it useful to characterize anisotropy of the DBOC by the absolute difference $|\gamma^{-1}-\gamma|$ that 
was inspired by the angular integral consideration in \eq{eq:t12}. For systems where the 
DBOC has cylindrical symmetry $|\gamma^{-1}-\gamma|=0$, while deviation from 
the cylindrical symmetry increases $|\gamma^{-1}-\gamma|$.

\begin{table}
  \caption{Parameters characterizing the importance of \gls{GP} effects in the studied systems. 
  The values have been obtained using 2D effective Hamiltonian parameters 
  and Eqs.~\eqref{eq:theta}, \eqref{eq:barw_ex}, and \eqref{eq:Cm_from_reff}.} 
  \label{tab:GP-param}
  \centering
  \begin{ruledtabular}
    \begin{tabular}{@{}lccr@{}}
      \multicolumn{1}{c}{$\gamma$} & $|\gamma^{-1} - \gamma|$ & $\bar w$, \% & $\bar w_{\rm app}$, \%  \\ \hline
      \multicolumn{4}{c}{ Bis(methylene) adamantyl cation} \\
      $0.09$ & $11.4$ & $42.2$ & $42.1$ \\[1ex]
      \multicolumn{4}{c}{ Butatriene cation} \\
      $0.67$ & $0.83$  & $87.8$ & $86.4$  \\[1ex]
      \multicolumn{4}{c}{ Pyrazine} \\
      $1.5$ & $0.82$ & $89.5$  & $73.5$
    \end{tabular}
  \end{ruledtabular}
\end{table}

\subsection{Bis(methylene) adamantyl cation}
\label{sec:bma-molecule}

\begin{figure}
  \centering
  \includegraphics[width=0.5\textwidth]{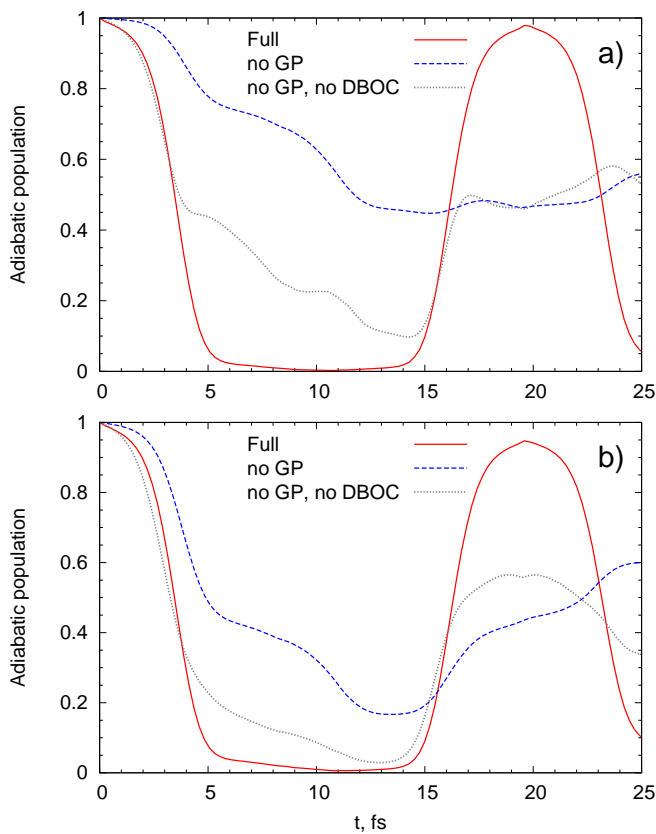}
  \caption{Excited state population dynamics of the \protect\gls{BMA} cation with different initial wave-packets:
  a) Gaussian wave-packet~\eqref{eq:Gaussian_wp}, b) the same as (a) but multiplied by the coupling coordinate.}
  \label{fig:bma}
\end{figure}

A high DBOC anisotropy  for the \gls{BMA} cation (Table~\ref{tab:GP-param})
suggests importance of GP effects via the \gls{DBOC} compensation mechanism. 
As evident from the adiabatic population dynamics for all
three models given in Fig.~\ref{fig:bma}a this is indeed the case.
The full model with the \gls{GP} demonstrates the fastest initial
population decay followed closely by the simplified model with no
\gls{GP} and no \gls{DBOC}. The model without the \gls{GP} (but with
the \gls{DBOC}) shows the slowest transfer since the wave-packet cannot
reach a strong coupling region due to the \gls{DBOC} repulsion.

Based on the average weight of the $m = 0$ component of the \gls{CWE}
[Eq.~\eqref{eq:wp-exp}]  $\bar w= 42.2\%$ (see Table~\ref{tab:GP-param}), 
GP modification of the $m = 0$ transfer can play a role in deviation of the simplified model dynamics from that of 
the full model. 
Comparing dynamics without the $m=0$ component (Fig.~\ref{fig:bma}b) shows that even in this 
highly anisotropic case removing the $m=0$ component visibly increases the overall transfer 
in the simplified model and makes it closer to the exact result.

 The frozen Gaussian estimate of the $m = 0$ component weight 
 $\bar w_{\rm app} = 42.1$\% is in excellent
agreement with the exact value $\bar w= 42.2\%$. Due to a high energy excess the
wave-packet slides quickly along the tuning coordinate toward the \gls{CI} point,
preserving the Gaussian form.  Thus, all assumptions made in the
derivation of Eq.~\eqref{eq:Cm_from_reff} are satisfied in this system.

A distinct feature of the full model is coherent oscillations
of the adiabatic population. They can be easily understood considering 
the dynamics in the diabatic representation where, due to very weak linear 
couplings, the initial wave-packet oscillates coherently on a single diabatic potential. 
These oscillations result in the oscillatory adiabatic population dynamics because 
regions of the diabatic potential before and after an intersection region correspond 
to different adiabatic states.
In the simplified (``no GP, no DBOC'') model the adiabatic population oscillations 
have similar frequency as in the full model but have quite different amplitude and more fine
structural elements. 
The first CI passage dynamics is very similar in both models 
but the difference increases when the wave-packet reflected by the repulsive part 
of the ground state potential returns to the CI point.  
On this returning trajectory absence of the DBOC in the simplified model 
allows the wave-packet not only to transfer back to the excited state but also to pass through
the \gls{CI} point remaining on the ground adiabatic surface. 
Thus, in the simplified model, the wave-packet \emph{bifurcates} at the \gls{CI}, 
and this bifurcation gives rise to nuclear decoherence that damps the coherent oscillations.
 
\subsection{Butatriene cation}
\label{sec:mathrmc_4h_4+}

\begin{figure}
  \centering
  \includegraphics[width=0.5\textwidth]{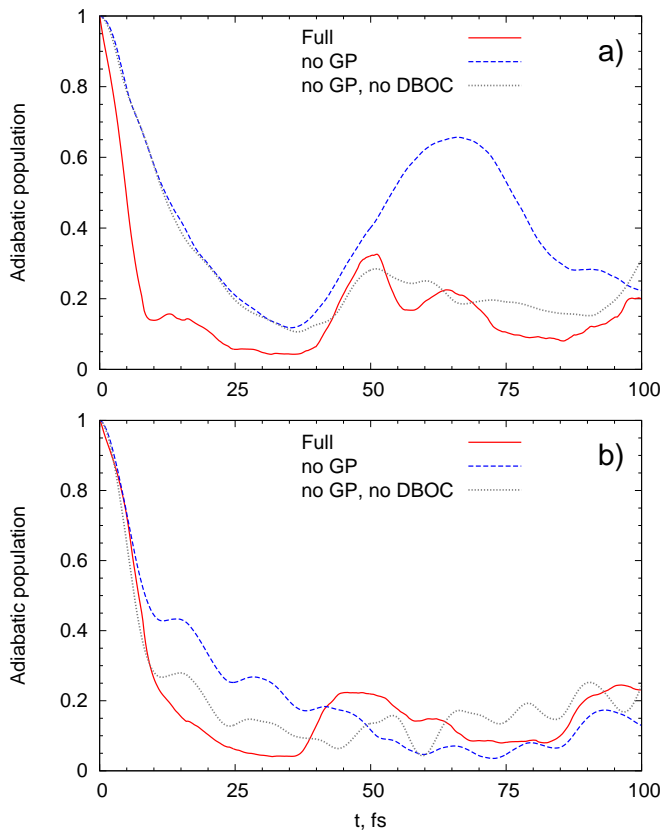}
  \caption{Excited state population dynamics of $\rm C_4H_4^{+}$ with different initial wave-packets:
  a) Gaussian wave-packet~\eqref{eq:Gaussian_wp}, b) the same as (a) but multiplied by the coupling coordinate.}
  \label{fig:c4h4}
\end{figure}

The butatriene cation has a relatively low anisotropy of the DBOC (see Table~\ref{tab:GP-param})
and is not expected to exhibit large \gls{GP} effects due to the DBOC compensation
mechanism. Indeed, short-time adiabatic population dynamics (Fig.~\ref{fig:c4h4}a)
shows almost no difference between models with and without the \gls{DBOC},
whereas dynamics in both models are quite different from that of the full model with the
\gls{GP}. The \gls{CWE} according to
Eq.~\eqref{eq:wp-exp} at the closest to the \gls{CI} position, shows
dominance of the $m = 0$ component with its average weight of $\bar w = 87.8$\%. 
Thus, as it is also seen from the dynamics with nodal initial Gaussian (Fig.~\ref{fig:c4h4}b),  
a role of the \gls{GP} for the butatriene cation is in facilitating transfer of the $m=0$ component. 
Figure~\ref{fig:c4h4}a illustrates that including the \gls{GP} 
can reduce an initial population transfer time-scale in 2-3 times with respect to 
those of models without the \gls{GP}.

The frozen Gaussian estimate of the $m=0$ component weight 
($\bar w_{\rm app} = 86.4\%$) is in excellent
agreement with the exact value due to spatial proximity of the initial 
\gls{FC} position of the wave-packet and the
\gls{CI} point.  The \gls{FC} point, which corresponds to
the ground-state minimum of the neutral molecule, is
located only 8.8 a.u. apart from the \gls{CI} point. 
The initial Gaussian distribution simply does not
have time to change its shape appreciably. 

\subsection{Pyrazine molecule}
\label{sec:pyrazine-molecule}

\begin{figure}
  \centering
  \includegraphics[width=0.5\textwidth]{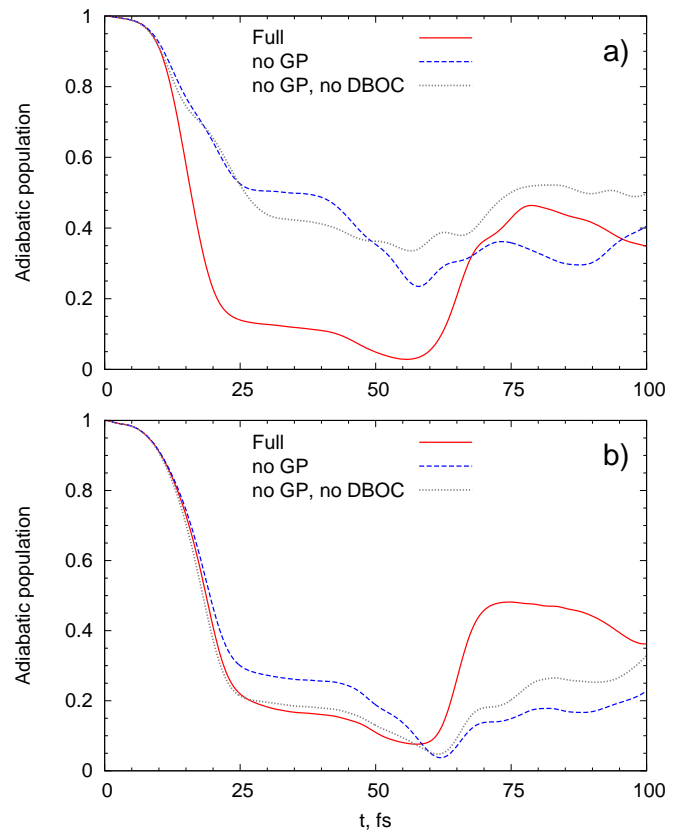}
  \caption{Excited state population dynamics of pyrazine with different initial wave-packets:
  a) Gaussian wave-packet~\eqref{eq:Gaussian_wp}, b) the same as (a) but multiplied by the coupling coordinate.}
  \label{fig:pyr}
\end{figure}

Comparing parameters for pyrazine and the butatriene cation in Table~\ref{tab:GP-param} 
we find surprising similarity that should result in similar dynamical trends: relative insignificance of 
the DBOC compensation and dominance of the $m=0$ transfer correction.
Indeed, the excited state adiabatic population dynamics given in
Fig.~\ref{fig:pyr}a
confirms that the \gls{DBOC} repulsion does \emph{not} contribute much
to the difference between models with and without the \gls{GP}.
Also, if we remove the $m=0$ component from the initial wave-packet 
the adiabatic populations of all three models become similar (see Fig.~\ref{fig:pyr}b).

The only small difference between the pyrazine molecule and butatriene cation according to Table~\ref{tab:GP-param}
is that the approximate weight of the $m=0$ component for pyrazine has somewhat poorer agreement with its exact value.
This deviation can be explained by a relatively long 48 a.u. spatial separation between the \gls{FC} and \gls{CI} points in pyrazine. 
The nuclear wave-packet does not go directly to 
the \gls{CI} point and spends substantial time in other regions of space, changing
the shape. Thus, the frozen Gaussian approximation is less accurate in this case.

\section{Concluding remarks}
\label{sec:conclusions}

Two cornerstones of our analysis of \gls{GP} effects in radiationless transitions of 
molecular systems through CIs are the transformation of the $N$-dimensional LVC model to
the effective 2D \gls{LVC} model, and the local analysis of the latter in 
the adiabatic representation. For the effective 2D Hamiltonian the \gls{GP} has been introduced by
transforming the Hamiltonian with the \citet{Mead:1979/jcp/2284} position-dependent phase factor.

Our local analysis revealed two main mechanisms of the \gls{GP} contribution to
non-adiabatic transitions. First, the \gls{GP} compensates for 
repulsion caused by the \gls{DBOC}, and second, it enhances transfer probability for
 a component of a nuclear wave-packet that corresponds to the zero eigenvalue 
 of the $L_z$ operator defined with respect to the CI point.

Two indicators have been introduced to quickly assess both \gls{GP} effects:
the anisotropy of the dimensionless coupling strength $|\gamma^{-1}-\gamma|$, 
and the weight $\bar w$ of the $m = 0$ component in cylindrical wave
expansion~\eqref{eq:wp-exp}. The former can be readily calculated
from parameters of the nuclear Hamiltonian, whereas the latter requires 
a dynamical simulation with the effective 2D Hamiltonian. 
Considering dynamics of a frozen Gaussian wave-packet 
with the assumption of orthogonality between coupling and tuning modes 
we have proposed the estimate of $\bar w$ [\eq{eq:Cm_from_reff}] that 
can be evaluated without dynamical simulations but using only 
a wave-packet width and the potential energy difference between 
an initial and CI points.

Using numerical simulations of adiabatic population dynamics 
for the \gls{BMA} and butatriene $\rm C_4H_4^{+}$ cations and the pyrazine molecule, it is
shown that the introduced indicators allow a reliable prediction of the \gls{GP} role
for studied systems. All systems exhibited substantial
\gls{GP} effects that can alter initial population transfer time-scales by factor of 2 to 3. 
Interestingly, GP effects in the studied systems modify non-adiabatic dynamics 
through different mechanisms. For the \gls{BMA} cation
the \gls{GP} compensates for the \gls{DBOC}
repulsion, and for the butatriene cation and pyrazine molecule it
strongly enhances non-adiabatic transition for the $m=0$ component of an 
incident wave-packet.

All systems treated in this paper were chosen so that the 
$N$-dimensional LVC model would be adequate for them. An 
interesting question is whether our treatment can be extended to 
more general non-LVC Hamiltonians. Since the core of our analysis 
is the local consideration in the vicinity of a CI point we can claim that as long 
as a nuclear wave-function approaches the CI seam close enough for 
the LVC parametrization to be accurate our analysis will be adequate. 
To confirm these ideas through numerical simulations we plan to apply the developed analysis to non-LVC 
models of CIs in pyrrole.\cite{Althorpe:2008/jcp/214117, Bouakline:2014/cp/}

Finally, in view of the DBOC compensating role of the GP
it is clear why common approximations omitting the DBOC and 
GP contributions work quite well together in 
mixed quantum-classical non-adiabatic dynamics simulations. In addition, 
for non-adiabatic dynamics near the CI, adding the DBOC term should be 
accompanied by including the GP. Adding only the DBOC term
without the GP in the best case will not affect dynamics appreciably but in 
the worst case can create uncompensated artificial repulsion and qualitatively 
incorrect dynamics. We hope that the proposed analysis will stimulate 
developments of new approximate methods for non-adiabatic dynamics 
in the adiabatic representation and will be of use in understanding 
results of simulations of non-adiabatic processes.

\section{Acknowledgments}

A.F.I. thanks Paul Brumer for stimulating discussions and acknowledges 
funding from the Natural Sciences and Engineering Research Council  
of Canada (NSERC) through the Discovery Grants Program.
L.J.D. is grateful to the European Union Seventh Framework Programme
(FP7/2007-2013) for the financial support under grant agreement PIOF-GA-2012-332233.

\appendix*

\section{Effective reduced dimensional model}
\label{sec:exactness-short-time}

Below we describe the transformation from the $N$-dimensional LVC Hamiltonian [Eq.~\eqref{eq:nd-lvc}] 
to the effective 2D Hamiltonian [\eq{eq:Halmost2D}], and show that the reduced model 
can reproduce the short time population dynamics of the full model. 

\subsection{Reduction procedure}

Recently, there has been significant progress in understanding how short time dynamics of the 
$N$-dimensional LVC model can be simulated using low dimensional Hamiltonians.
\cite{Cederbaum:2005/prl/113003, Burghardt:2006/mp/1081,Gindensperger:2006/jcp/144103} 
Cederbaum and coworkers\cite{Burghardt:2006/mp/1081} have shown several approaches to building low dimensional 
effective Hamiltonians that reproduce short-time dynamic of the full Hamiltonian. The reason for this 
success was found comparing cumulant expansions of the auto-correlation functions of the effective and full Hamiltonians.
With only three effective modes it is possible to construct an effective Hamiltonian that will reproduce three first 
cumulants of the total Hamiltonian. In our previous work on GP effects for low energy dynamics 
we developed a transformation similar to those proposed by Cederbaum and coworkers 
with the crucial difference that our transformation resulted in only a two dimensional subsystem.\cite{Loic:2013/jcp/234103}  
In the current work the dynamical properties of our transformation have been improved
by introducing a frequency weighting step. As shown below, this step creates better agreement 
between time derivatives of electronic population dynamics for the effective and full Hamiltonians. 

\paragraph{Frequency weighting.---}
Starting with the $N$-dimensional \gls{LVC} Hamiltonian [Eq.~\eqref{eq:nd-lvc}]
we modify its coordinates $\tilde q_j = \sqrt{\omega_j}q_j$ 
and momenta $\tilde p_j = p_j/\sqrt{\omega_j}$. 
The resulting Hamiltonian is
\begin{align}
  \label{eq:app-H1}
   H_1 = {} & \left[\frac{1}{2}\left(\tilde{\mathbf p}^\dagger{\boldsymbol\omega}\tilde{\mathbf p} + \tilde{\mathbf q}^\dagger{\boldsymbol\omega}\tilde{\mathbf q} \right) + \tilde{\mathbf f}^\dagger\tilde{\mathbf q}\right] \mathbf{1}_2 \nonumber \\
  & {} +
  \begin{pmatrix}
    \tilde{\mathbf d}^\dagger\tilde{\mathbf q} &  \tilde{\mathbf c}^\dagger\tilde{\mathbf q} \\ 
    \tilde{\mathbf c}^\dagger\tilde{\mathbf q} & -\tilde{\mathbf d}^\dagger\tilde{\mathbf q}
  \end{pmatrix} +
  \begin{pmatrix}
    \delta/2 & 0 \\
    0 & -\delta/2
  \end{pmatrix},
\end{align}
where a vector notation is introduced: $\tilde{\mathbf q} = \{\tilde q_j\}_{j=1}^N$, $\tilde{\mathbf p} = \{\tilde p_j\}_{j=1}^N$, $\tilde{\mathbf d} = \{\frac{\kappa_j-\tilde\kappa_j}{2\sqrt{\omega_j}}\}_{j=1}^N$, $\tilde{\mathbf f} = \{\frac{\kappa_j+\tilde\kappa_j}{2\sqrt{\omega_j}}\}_{j=1}^N$, $\tilde{\mathbf c} = \{c_j/\sqrt{\omega_j}\}_{j=1}^N$, and ${\boldsymbol\omega} = \diag{\{\omega_1,\ldots, \omega_N\}}$ is a diagonal matrix of frequencies.

\paragraph{Definition of the effective coordinates.---} To perform a truncation that would keep 
all non-adiabatic effects within a two-dimensional subspace 
we define a new set of coordinates $\{\tilde Q_1,\tilde Q_2,\hdots, \tilde Q_N\}$ obtained 
from $\{\tilde q_1,\tilde q_2,\hdots,\tilde q_N\}$ by an orthogonal transformation $\mathbf{O}_1$:
$\tilde{\mathbf Q}={\mathbf O}_1\tilde{\mathbf q}$.
The first two rows of ${\mathbf O}_1$ define a 2D subsystem of the 
effective coordinates $\tilde Q_1$ and $\tilde Q_2$ 
\begin{equation}
  \label{eq:app-O1}
  \begin{pmatrix}
    {\mathbf e}^T_d\\
    (\tilde{\mathbf c}^T-\tilde C_1 {\mathbf e}^T_d)/\tilde C_2
  \end{pmatrix},
\end{equation}
where
\begin{align}
  {\mathbf e}_d {} =& \tilde{\mathbf d}/||\tilde{\mathbf d}||, \nonumber\\
    \tilde C_1 {} =& \tilde{\mathbf c}\cdot{\mathbf e}_d, \nonumber\\
    \tilde C_2 {} =& \sqrt{||\tilde{\mathbf c}||^2 - (\tilde{\mathbf c}\cdot{\mathbf e}_d)^2}.
\end{align}
The remainder of ${\mathbf O}_1$ and coordinates $\{\tilde Q_j\}_{j=3}^N$ are defined 
by  the Gram-Schmidt orthogonalization procedure with respect to the $\{\tilde Q_1,\tilde Q_2\}$  subspace. 
In the $\{\tilde Q_i\}$ representation the Hamiltonian becomes
\begin{align}
  \label{eq:app-H2}
  H_2 = & \left[ \frac{1}{2}\left(  \tilde{\mathbf P}^\dagger{\boldsymbol\Lambda}\tilde{\mathbf P} + \tilde{\mathbf Q}^\dagger{\boldsymbol\Lambda}\tilde{\mathbf Q}\right) + \tilde{\mathbf F}^\dagger\tilde{\mathbf Q}\right]{\mathbf 1}_2
+\begin{pmatrix}
     \delta/2 & 0 \\
     0 & -\delta/2
  \end{pmatrix} \nonumber\\
&+\begin{pmatrix}
    \tilde D_1\tilde Q_1 & \tilde C_1\tilde Q_1 +  \tilde C_2\tilde Q_2 \\
    \tilde C_1\tilde Q_1 + \tilde C_2\tilde Q_2 & -\tilde D_1\tilde Q_1
  \end{pmatrix},
\end{align}
where ${\boldsymbol\Lambda}={\mathbf O}_1{\boldsymbol\omega}{\mathbf O}_1^\dagger$, 
$\tilde{\mathbf F}={\mathbf O}_1\tilde{\mathbf f}$, and $\tilde D_1 = ||\tilde{\mathbf d}||$.
A convenient feature of the $H_2$ Hamiltonian is that all differences between electronic 
surfaces and couplings are concentrated in the two-dimensional $\{\tilde Q_1,\tilde Q_2\}$  subspace.
Next, we truncate the full set of 
coordinates $\tilde{\mathbf Q}_\text{S}=\left(\begin{smallmatrix}\tilde Q_1\\\tilde Q_2\end{smallmatrix}\right)={\boldsymbol\Pi}\tilde{\mathbf Q}$ and momenta $\tilde{\mathbf P}_\text{S}=\left(\begin{smallmatrix}\tilde P_1\\\tilde P_2\end{smallmatrix}\right)={\boldsymbol\Pi}\tilde{\mathbf P}$ to 
the two-dimensional subspace using a projector ${\boldsymbol\Pi}$. 
This truncation leads to the two-dimensional effective Hamiltonian
\begin{align}
  \label{eq:app-H3}
  H_3 = & \left[ \frac{1}{2}\left( \tilde{\mathbf P}_\text{S}^\dagger{\boldsymbol\Lambda}_\text{S}\tilde{\mathbf P}_\text{S} + \tilde{\mathbf Q}_\text{S}^\dagger{\boldsymbol\Lambda}_\text{S}\tilde{\mathbf Q}_\text{S}\right) + \tilde{\mathbf F}_\text{S}^\dagger\tilde{\mathbf Q}_\text{S}\right]{\mathbf 1}_2 \nonumber\\
&{}+\begin{pmatrix}
    \tilde{\mathbf D}_\text{S}^\dagger\tilde{\mathbf Q}_\text{S} & \tilde{\mathbf C}_\text{S}^\dagger\tilde{\mathbf Q}_\text{S} \\
    \tilde{\mathbf C}_\text{S}^\dagger\tilde{\mathbf Q}_\text{S} & -\tilde{\mathbf D}_\text{S}^\dagger\tilde{\mathbf Q}_\text{S}
  \end{pmatrix} +
  \begin{pmatrix}
     \delta/2 & 0 \\
     0 & -\delta/2
  \end{pmatrix}.
\end{align}
All vectors and matrices are assigned the subscript S to indicate their two-dimensional character.

\paragraph{Extra transformations to the 2D \gls{LVC} Hamiltonian.---}
To arrive at a subsystem Hamiltonian that is closer in form to the 2D LVC Hamiltonian [\eq{eq:H_lvc}] 
 we diagonalize the frequency matrix ${\boldsymbol\Lambda}_\text{S}$ 
with the orthogonal transformation 
${\mathbf O}_2$, ${\boldsymbol\Omega}={\mathbf O}_2{\boldsymbol\Lambda}_\text{S}{\mathbf O}_2^\dagger$, 
and reverse the frequency weighting of coordinates.
These transformations lead to new effective coordinates 
${\mathbf Q}={\boldsymbol\Omega}^{-\frac{1}{2}}{\mathbf O}_2\tilde{\mathbf Q}_\text{S}$, 
momenta ${\mathbf P}={\boldsymbol\Omega}^{\frac{1}{2}}{\mathbf O}_2\tilde{\mathbf P}_\text{S}$, and the 
Hamiltonian
\begin{align}
  \label{eq:app-H4}
  H_4 = & \left[ \frac{1}{2}\left( {\mathbf P}^\dagger{\mathbf P} + {\mathbf Q}^\dagger{\boldsymbol\Omega}^2{\mathbf Q}\right) + {\mathbf F}^\dagger{\mathbf Q}\right]{\mathbf 1}_2 \nonumber\\
&{}+\begin{pmatrix}
    {\mathbf D}^\dagger{\mathbf Q} & {\mathbf C}^\dagger{\mathbf Q} \\
    {\mathbf C}^\dagger{\mathbf Q} & -{\mathbf D}^\dagger{\mathbf Q}
  \end{pmatrix} +
  \begin{pmatrix}
     \delta/2 & 0 \\
     0 & -\delta/2
  \end{pmatrix},
\end{align}
where ${\mathbf F}={\boldsymbol\Omega}^{\frac{1}{2}}{\mathbf O}_2\tilde{\mathbf F}_\text{S}$, 
${\mathbf D}={\boldsymbol\Omega}^{\frac{1}{2}}{\mathbf O}_2\tilde{\mathbf D}_\text{S}$, 
${\mathbf C}={\boldsymbol\Omega}^{\frac{1}{2}}{\mathbf O}_2\tilde{\mathbf C}_\text{S}$.
Finally, we translate the origin of the the 2D subspace ${ X} = {Q}_1 + {\Omega_1}^{-2}{ F}_1$, 
${ Y} = {Q}_2 + {\Omega_2}^{-2}{ F}_2$ 
and obtain the Hamiltonian given in \eq{eq:Halmost2D}
\begin{align}
  \label{eq:app-H5}
   H_{2D} = {} & \left(\frac{P_X^2 + P_Y^2}{2} + \frac{{
        \Omega}_1^2 X^2 + {\Omega}_2^2 Y^2}{2}\right) \mathbf{1}_2 +
  \begin{pmatrix}
    \frac{\Delta}{2} & \Delta_{12} \\
    \Delta_{12} & -\frac{\Delta}{2}
  \end{pmatrix}\nonumber\\ & {} +
  \begin{pmatrix}
    D_1 X + D_2 Y &  C_1 X + C_2 Y \\
    C_1 X + C_2 Y & -D_1 X - D_2 Y
  \end{pmatrix},
\end{align}
where $\Delta=\delta-2{\mathbf D}^\dagger{\boldsymbol\Omega}^{-2}{\mathbf F}$, and
$\Delta_{12}=-{\mathbf C}^\dagger{\boldsymbol\Omega}^{-2}{\mathbf F}$. Note that the \gls{FC} point, 
 initially at the origin of the coordinate system in the $N$-dimensional space, 
is now shifted by the vector ${\boldsymbol\Omega}^{-2}{\mathbf F}$.

\subsection{Short-time population dynamics}

To assess the difference in short time population dynamics for the full [\eq{eq:app-H2}] 
and reduced [\eq{eq:app-H3}] models we compare low order terms of 
population Taylor time series for both models. However, due to a non-polynomial form 
of the adiabatic Hamiltonian [\eq{eq:adiab}], 
derivation of analytical expressions for adiabatic populations becomes intractable. 
To avoid this complication we focus on the \emph{diabatic} population $P_\text{dia}(t)$:
\begin{equation}
  \label{eq:app-Pdia}
  P_\text{dia}(t) =
  \braket{\Psi(0)|e^{i\hat H_\text{dia}t}{\hat P}_\text{dia}e^{-i\hat H_\text{dia}t}|\Psi(0)}.
\end{equation}
Here, ${\hat P}_\text{dia}=\left(\begin{smallmatrix}1&0\\0&0\end{smallmatrix}\right)$ 
is the projector to the diabatic state that has higher energy in the FC point,
$\hat H_\text{dia}$ is a general diabatic Hamiltonian that can be either $H_2$ or $H_3$, and
$\ket{\Psi(0)}$ is the initial total wave-function. 
We expand $P_\text{dia}(t)$~\eqref{eq:app-Pdia} in a Taylor series
\begin{align}
  \label{eq:expandP}
  P_\text{dia}(t) = & \sum_{k=0}^{\infty} \frac{t^k}{k!} M_k
\end{align}
where $M_k = d^k P_\text{dia}(t)/dt^k\vert_{t=0}$. 
First few terms of this expansion define short-time dynamics and for the reduced model 
to reproduce the full model dynamics, corresponding terms of two expansions should 
be close. 
Using \eq{eq:app-Pdia} $M_k$'s can be alternatively defined as
\begin{align}
  \label{eq:momemts}
  M_k = & i^k \sum_{l=0}^{l=k} \frac{(-1)^l k!}{l!(k-l)!}
  \braket{\Psi(0)|\hat H_\text{dia}^{k-l}{\hat P}_\text{dia}\hat H_\text{dia}^l|\Psi(0)}.
\end{align}
Due to time reversal symmetry of the population dynamics at $t=0$ all odd derivatives are zero. 
Using $N$-dimensional Gaussian wave-packet in the initial conditions 
[$\Psi(0)\propto (\exp\left\{-\tilde{\mathbf Q}^\dagger\tilde{\mathbf Q}/2\right\},0)^{\dagger}$]
and Gaussian integration the first three even orders of $M_k$ 
for the $N$-dimensional Hamiltonian (\ref{eq:app-H2}) are obtained
\begin{align}\label{eq:app-M0}
  M_0 = {} &  1,\\\label{eq:app-M2}
  M_2 = {} & -\tilde{\mathbf C}^\dagger\tilde{\mathbf C},\\\label{eq:app-M4}
  M_4 = {} &  \tilde{\mathbf C}^\dagger\left( {\boldsymbol\Lambda} - \delta{\mathbf 1}_N \right)^2\tilde{\mathbf C} + 4 \left(\tilde{\mathbf C}^\dagger\tilde{\mathbf D}\right)^2\\
  {} & + \tilde{\mathbf C}^\dagger\tilde{\mathbf C} \left( 6 \tilde{\mathbf C}^\dagger\tilde{\mathbf C} + 2 \tilde{\mathbf D}^\dagger\tilde{\mathbf D}\right).\nonumber
\end{align}
Corresponding terms for the reduced model [\eq{eq:app-H3}] are evaluated similarly 
using the 2D Gaussian wave-packet 
$\Psi(0)\propto (\exp\left\{-\tilde{\mathbf Q}_S^\dagger\tilde{\mathbf Q}_S/2\right\},0)^{\dagger}$  
\begin{align}\label{eq:app-MS}
  M_{0,S} = {} &  1,\\\label{eq:app-M2s}
  M_{2,S} = {} & -\tilde{\mathbf C}_S^\dagger\tilde{\mathbf C}_S,\\\label{eq:app-M4s}
  M_{4,S} = {} &  \tilde{\mathbf C}_S^\dagger\left( {\boldsymbol\Lambda}_S - \delta{\mathbf 1}_2 \right)^2\tilde{\mathbf C}_S + 4 \left(\tilde{\mathbf C}_S^\dagger\tilde{\mathbf D}_S\right)^2\\
  {} & + \tilde{\mathbf C}_S^\dagger\tilde{\mathbf C}_S \left( 6 \tilde{\mathbf C}_S^\dagger\tilde{\mathbf C}_S + 2 \tilde{\mathbf D}_S^\dagger\tilde{\mathbf D}_S\right).\nonumber
\end{align}
The zeroth-order terms are the same in both expansions, 
while relations between the corresponding second- and fourth-order terms need some elaboration. 
By construction of the orthogonal transformation ${\mathbf O}_1$ [\eq{eq:app-O1}], 
$\tilde{\mathbf D}$ and $\tilde{\mathbf C}$ have $N-2$ zero entries: $\tilde C_j = \tilde D_j = 0$, $j = 3,\ldots, N$. 
Therefore, we have the following identity
\begin{align}
\tilde{\mathbf C}^\dagger\tilde{\mathbf C} = {} & \tilde{\mathbf C}^\dagger{\boldsymbol\Pi}\tilde{\mathbf C} = \tilde{\mathbf C}_\text{S}^\dagger\tilde{\mathbf C}_\text{S}
\end{align}
which proves that $M_{2,S}=M_2$. 
Similarly, all terms of $M_4$ but $\tilde{\mathbf C}^\dagger{\boldsymbol\Lambda}^2\tilde{\mathbf C}$ 
coincide with corresponding terms of $M_{4,S}$.
Generally we have
\begin{align}
  \tilde{\mathbf C}^\dagger{\boldsymbol\Lambda}^2\tilde{\mathbf C} = {} & \tilde{\mathbf C}^\dagger{\boldsymbol\Pi}{\boldsymbol\Lambda}^2{\boldsymbol\Pi}\tilde{\mathbf C} = \tilde{\mathbf C}_\text{S}^\dagger{\boldsymbol\Lambda}^2\tilde{\mathbf C}_\text{S} \nonumber\\ \label{eq:app-ineq}
{}  \neq {} & {} \tilde{\mathbf C}_\text{S}^\dagger{\boldsymbol\Lambda}_\text{S}^2\tilde{\mathbf C}_\text{S} = \tilde{\mathbf C}_\text{S}^\dagger({\boldsymbol\Pi}{\boldsymbol\Lambda}{\boldsymbol\Pi})^2\tilde{\mathbf C}_\text{S},
\end{align}
where the inequality is due to the existence of couplings 
between the two-dimensional subspace and the complementary space in ${\boldsymbol\Lambda}$. 
In the \gls{BMA} case, the relative error in $M_4$ due to the inequality 
turns out to be small $(M_4-M_{4,S})/M_4=8\cdot10^{-3}$. Moreover, 
for the butatriene cation and the pyrazine molecule, because the $\tilde{\mathbf c}$ vector in 
Eq.~\eqref{eq:app-H1} contains only a single non-zero component
the transformations from $H_1$ to $H_3$ leave $\tilde{\mathbf C}_\text{S}$
decoupled from the other coordinates. Therefore, for these systems, the inequality in
Eq.~\eqref{eq:app-ineq} becomes an equality and $M_4=M_{4,S}$. 

\begin{figure*}
  \centering
  \includegraphics[width=\textwidth]{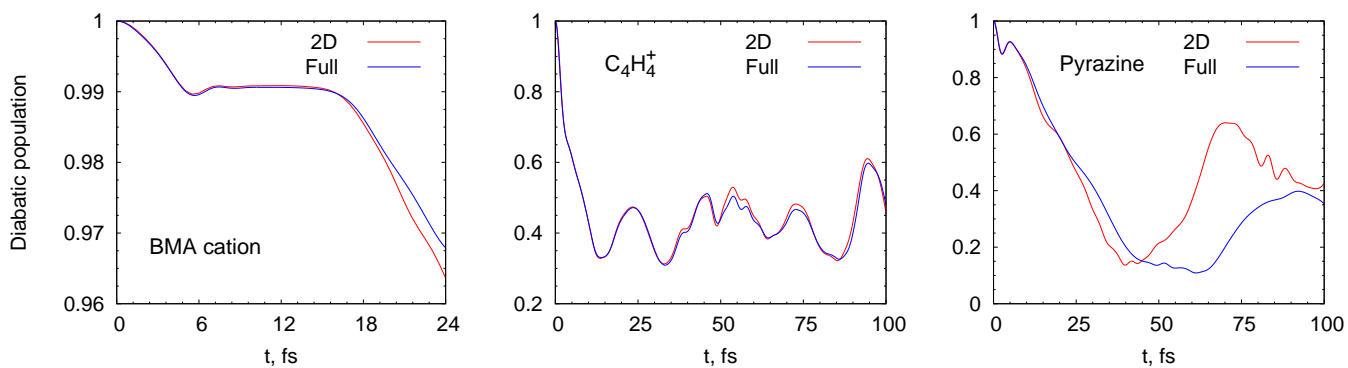}
  \caption{Diabatic population dynamics for the full and effective two-dimensional models.}
  \label{fig:comp-2D-ND}
\end{figure*}

\begin{figure*}
  \centering
  \includegraphics[width=\textwidth]{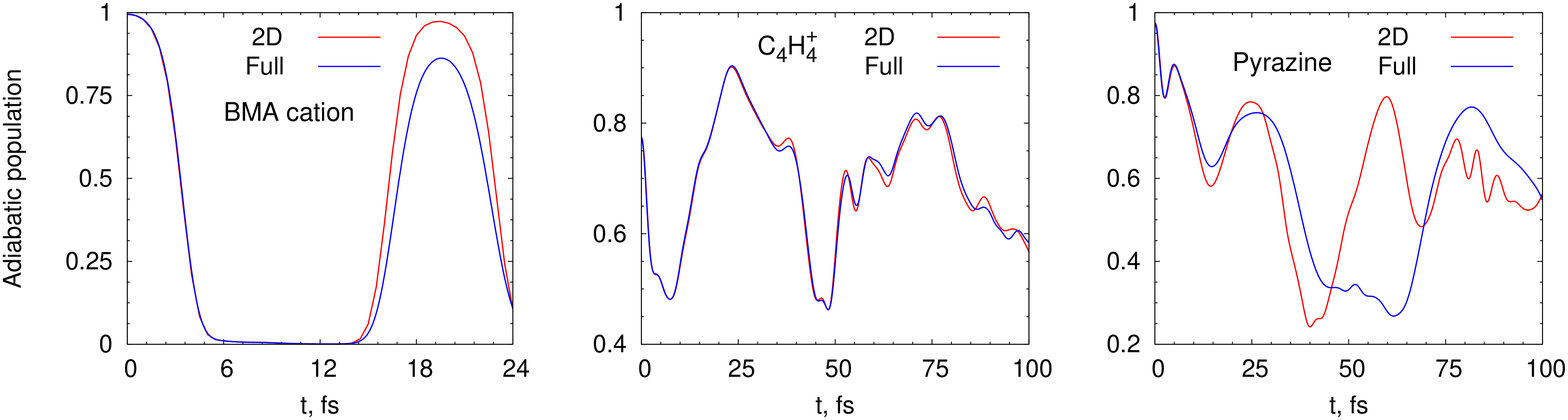}
  \caption{Adiabatic population dynamics for the full and effective two-dimensional models.}
  \label{fig:comp-2D-ND_adi}
\end{figure*}

Comparison of diabatic populations obtained with the full and reduced models are shown in Fig.~\ref{fig:comp-2D-ND}. 
To simulate dynamics in the full dimensional diabatic models we used the MCTDH package.\cite{mctdh}  
For all systems there is a very good agreement between results of the full and reduced model 
dynamics until the end of the first CI passage: $15$ fs for \gls{BMA} (see Fig.~\ref{fig:bma}), $35$ fs for 
the butatriene cation (see Fig.~\ref{fig:c4h4}), and $45$ fs for the pyrazine molecule 
(see Fig.~\ref{fig:pyr}). Similar trends we see for the adiabatic populations in Fig.~\ref{fig:comp-2D-ND_adi}, 
and thus we can conclude that the mode reducing transformation preserves the short-time non-adiabatics
dynamics in the studied systems very well.


%

\end{document}